\documentclass{aa}

\usepackage{graphics}
\newcommand{\rb}[1]{\raisebox{1.5ex}[0pt]{#1}}
\def\lsim{\,\lower2truept\hbox{${<\atop\hbox{\raise4truept\hbox{$\sim$}}}$}\,}
\def\gsim{\,\lower2truept\hbox{${>\atop\hbox{\raise4truept\hbox{$\sim$}}}$}\,}
\def\proptosim{\,\lower2truept\hbox{${\propto\atop\hbox{\raise4truept\hbox{$\i
m$}}}$}\,}

\begin{document}
\title{In-flight main beam reconstruction for Planck-LFI}
 
\subtitle{}
\author{C. Burigana$^1$ \and P. Natoli$^2$ \and N. Vittorio$^2$ \and N. 
Mandolesi$^1$
\and M. Bersanelli$^3$ }

\institute{Istituto TeSRE/CNR, via P.~Gobetti 101, I-40122 Bologna, Italy
\and Dipartimento di Fisica, Universit\`a di Roma ``Tor Vergata'', 
via della Ricerca Scientifica 1, I-00133, Roma, Italy
\and Dipartimento di Fisica, Universit\`a di Milano, and IFC/CNR, 
via Celoria 16, I-20133, Milano, Italy}

\offprints{burigana@tesre.bo.cnr.it}

\date{Submitted on A\&A, 02.11.2000}
\thesaurus{12(12.03.1; 12.04.2) 03(03.19.2; 03.20.9; 03.13.2) 07(07.19.2)}


\maketitle
\markboth{C. Burigana et~al.: In-flight main beam reconstruction for {\sc Planck}-LFI}
{C. Burigana et~al.: In-flight main beam reconstruction for {\sc Planck}-LFI}
\begin{abstract}

In-flight measurement of the antenna main beams of the {\sc Planck}
instruments is a crucial input to the data analysis pipeline.
We study the main beam reconstruction achievable through external planets 
using a flight simulator to model their observation.
We restrict our analysis to the 30~GHz LFI channel but the method can be 
easily extended to higher frequency channels.
We show that it is possible to fit the antenna response 
from Jupiter and Saturn to obtain an
accurate, robust, simple and fast reconstruction 
of the main beam properties under very general conditions,
independently of the calibration accuracy.
In addition, we find that a bivariate Gaussian approximation
of the main beam shapes represents a significant improvement with respect 
to a symmetric representation.
We also show that it is possible
to combine the detection of the planet's transit and 
Planck's very accurate in-flight calibration to measure 
the planet's temperature at millimetric wavelengths with an
accuracy at the \% level.
This work is based on {\sc Planck}-LFI activities.


\end{abstract}

\section{Introduction}
\label{intro}

The {\sc Planck} 
Surveyor\footnote{http://astro.estec.esa.nl/SA-general/Projects/Planck/} 
is the ESA space mission devoted to the study of the Cosmic Microwave
Background. {\sc Planck} will have an impact on a number of scientific
issues, such as the physics of the early universe, structure formation
theory and cosmological parameters determination (Bersanelli et al.\, 
1996). In order to reach the necessary level of sensitivity it is
important to understand systematics and to keep them under control. In this paper
we will focus on the behavior of the {\sc Planck} Low Frequency Instrument
(LFI, Mandolesi et~al. 1998) 
antenna patterns. For simplicity, we will restrict our analysis to the 30
GHz LFI channel but the method we present here can be easily extended to
higher frequency channels. 

The beam pattern is affected by optical distortions, which depend on the
telescope design and on the arrangement of the various feed horns in the
focal plane. These effects degrade both angular resolution and 
sensitivity  (e.g., Mandolesi et al.~2000a,b). Therefore,
accurate measurement of the beam pattern is a crucial input to the
data analysis pipeline.

Due to their small angular size external planets  produce large signals
only when seen in the main beam.  As such, they represent a unique
possibility to recover directly from the data  the in-flight behavior of
the main beam. This point was already addressed, in the framework of the
{\sc Planck} mission, by Bersanelli et~al. 1997 for the simple case of
a Gaussian symmetric antenna response.  We extend here this analysis to
quantify our ability to reconstruct   a more realistic, asymmetric beam
pattern. 

The plan of this paper is as follows. In Sect.~2 we describe our main
tools and assumptions. In Sect.~3 we discuss the quality of the main beam
reconstruction and the implications for the treatment of {\sc Planck} data. 
As a byproduct of our simulations, we also verify 
in Sect.~\ref{internal_straylight}
the validity of the {\sc Planck} optical design to strongly suppress  
straylight contamination from internal Solar System bodies.
In Sect.~5 we summarize our findings and draw our conclusions.
\section{Method}
\label{simul}

In order to attack the problem of the in-flight main beam reconstruction,  we  have
to: (\textit{i}) describe the {\sc Planck} orbit and scanning strategy; 
(\textit{ii}) quantify the antenna response;
(\textit{iii}) exploit the planet's mm emission  and positions;
(\textit{iv}) simulate the {\sc Planck} observations of the external planets
(basically Jupiter and Saturn).
Here we briefly discuss these points separately.

\subsection{{\sc Planck} orbit and scanning strategy}

The selected orbit for the {\sc Planck} satellite is a Lissajous orbit  around the
Lagrangian point L2 of the Sun-Earth system (e.g., Bersanelli et~al.  1996). 
In the nominal operation scheme the
spacecraft spins at 1 r.p.m. around an axis 
kept parallel to the ecliptic plane. Every
hour the spin axis is moved by $2.5'$ maintaining its anti-solar direction. The
telescope optical axis is at an angle $\alpha$ from the spin axis direction.  
 The
spin axis might precede about the  anti-solar direction, with a period of about
six months and an amplitude of about $10^\circ$. This spacecraft movement is of course 
over imposed to the Lissajous orbit and to the spin axis hourly  shift. In this paper we consider 
values of $\alpha$ between $80^\circ$ and $90^\circ$, i.e. about the value of $85^\circ$ 
recently recommended by the {\sc Planck} Science Team.  We make use of the
{\sc Planck} flight simulator described in detail by Burigana et~al. 1997, 1998
and Maino et~al. 1999 properly modified to model
the {\sc Planck} observations of the Solar System bodies and the spacecraft
motion (see, e.g., Bersanelli et~al. 1997). 

For what follows, it is convenient to introduce a telescope ``reference frame''
(hereafter rf) 
$\{x_T,y_T,z_T\}$ with the
$z_T$ axis coincident with the direction of the telescope line of sight
(the ${\hat p}$ direction, say) and with the $x_T-y_T$ plane 
identifying the telescope field of view plane (we choose to
orient  ${\hat x}_T$  towards the intersection of the $x_T-y_T$ plane 
with the spin axis ${\hat s}$ or, in the case $\alpha = 90^\circ$, 
${\hat x}_T \parallel {\hat s}$). 
For the considered scanning strategies the spin axis and the
telescope directions (${\hat s}$ and ${\hat p}$, respectively) are easily derived given
the observation time, the spinning frequency and the boresight angle $\alpha$. So it
is always possible to pass from a chosen celestial rf to the telescope rf (and viceversa) by a
suitable Eulerian rotation of the considered rf.

\subsection{Antenna angular pattern}

The {\sc Planck} High Frequency Instrument (HFI, Puget et~al. 1998) is located 
at the center of the focal plane. 
The LFI feed horns surround HFI and  are then substantially off-axis. For
instance, with a telescope of $1.5$~m class,
the 30~GHz beams are at about
$\simeq 5^{\circ}$ from ${\hat p}$. So,   it is convenient to define a beam rf
$\{x_b,y_b,z_b\}$ with the
$z_b$ axis coincident with  beam axis $\hat b$ and with the  
 $x_b-y_b$ plane slightly tilted with respect to the telescope field of view plane 
(we keep the convention of obtaining the beam rf from the telescope rf through 
a rotation of the telescope rf about an axis hortogonal 
to the plane identified by ${\hat p}$ and ${\hat b}$ 
by the angle necessary to transport ${\hat p}$ to ${\hat b}$.

As we know the position of each feed horn
in the focal plane, we
can always pass from the telescope rf to  the beam rf and viceversa. In fact,  the flight
simulator determines for every time step  the orientations in the sky of the telescope 
and beam reference frames,  to compute the  antenna response for a given line of sight.

\subsubsection{Main beam}
\label{coord}

Recent improvements on the {\sc Planck} optical
design based on aplanatic solutions (Mandolesi et~al. 2000b) show that the main
beams are roughly elliptical,  with an ellipticity ratio $r \lsim 1.4$ 
(Alcatel, private reference, PL-AS-TN-022).
Therefore, we approximate the antenna pattern as an off-axis
bivariate Gaussian beam. For simplicity, we will consider the bivariate Gaussian
beam projected onto the field of view plane (i.e. we will consider this beam 
representation on the $x_T-y_T$ plane, and not on the 
$x_b-y_b$ beam plane). 
To be accurate one has to say that if the true beam shape is elliptical 
in the $x_b-y_b$ plane, it gets
distorted by the projection on the $x_T-y_T$ plane. 
However,  since the off-axis angle even for the 30 GHz beam is
small,  this distortion is negligible  and, if anything, 
does not change the elliptical nature
of the beam response. In addition, a realistic main beam distortion 
implies a deviation from the elliptical shape larger 
than that introduced by this projection.
So, let $(x^*_T,y^*_T)$ identify the projection of the beam centre unit vector
onto the
$x_T-y_T$ plane. Let
$\epsilon$ be the angle between the $x_T$-axis and the principal axis of the bivariate
Gaussian. The (normalized to the maximum) beam response can be then 
expressed as:  
\begin{equation}\label{par_beam} J=\exp[-\frac{1}{2}\;(\mathcal{R}\mathbf{u})^t
\Sigma^{-1}\mathcal{R}\mathbf{u}] 
\end{equation} where
\[
\mathbf{u}=\left( \begin{array}{c} x_T-x^*_T \\ y_T-y^*_T \end{array}\right)\;,
\]
$\Sigma=\mathrm{diag}(\sigma^2_+,\sigma^2_-)$ contains the bivariate's
 beam dispersions along the ellipse principal axis and $\mathcal{R}$ is the rotation 
matrix for an angle
$\epsilon$ in the $x_T-y_T$ plane:
\[
\mathcal{R}=\left( \begin{array}{cc} \cos\epsilon \; \sin\epsilon \\
                         -\sin\epsilon \; \cos\epsilon \end{array} \right).
\] It is then convenient to define the beam ``sigma'' $\sigma = \sqrt{\sigma_+ 
\sigma_-}$ and the ellipticity ratio $r = \sigma_+ / \sigma_- $. 

\subsubsection{Far side lobes}

A realistic
description of the antenna far side lobes has to rest on the accurate optical
calculations of de Maagt et~al. 1998 for the {\sc Planck} telescope including
shields. 
Several cuts at constant azimuthal angle $\phi$ 
are shown in Fig.~1 
as a function of the colatitude
angle $\theta$ from the beam axis ${\hat b}$. 
In the ``antispillover'' region 
[about $(\theta , \phi) \sim (90^\circ , 180^\circ)$]  
where it is mostly important to evaluate the effects of straylight from the
internal bodies of the Solar System (i.e. Sun, Earth and Moon)
the pattern response drops down to approximately $\sim 100 \rm{dB}$. 
At such a level of
rejection, according to  the {\sc Planck} requirements, one expects 
that Sun, Earth and Moon are completely harmless for the mission. 
We will further discuss this point in Sect.~\ref{internal_straylight}.

\begin{figure}
\resizebox{\hsize}{!}{\includegraphics{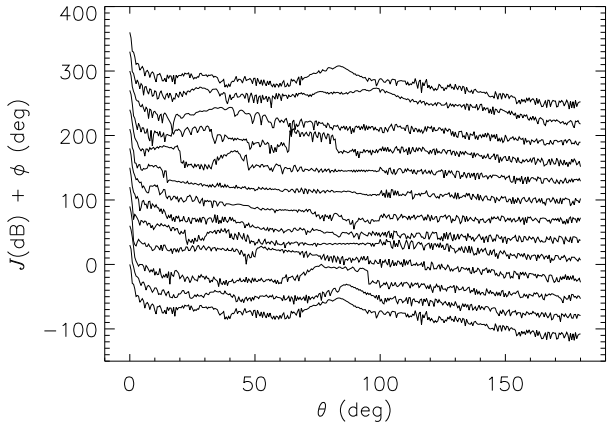}} 
\caption{Several cuts of the full antenna pattern, in dB, 
as computed by de Maagt et~al. 1998.
The lines refer to the antenna response for $\theta$
between $0^{\circ}$ and $180^{\circ}$ and, from the bottom to the top, with
$\phi$ at steps of $30^{\circ}$ from $0^{\circ}$ to $360^{\circ}$
 (each cut is vertically
shifted for graphic purposes to have the value at the point at 
$\theta=0^{\circ}$ equal to the considered value of $\phi$).}
\label{full_patt_cuts}
\end{figure}

\subsection{Planet's mm emissions and positions }
\label{sol_sys}
\subsubsection{External Planets}
Several authors reported measurements of the  planets brightness temperature at millimeter 
wavelengths with typical uncertainties of $3 \div 5\%$ (see, e.g., 
Bersanelli et~al. 1997 and references therein). 
The quite large uncertainties associated with these values  prevents one
from using planets for accurate temperature calibration of
the {\sc Planck} time order data.  This will be done, to better  than a
$1\%$,   by using the diffuse signature of the CMB dipole  anisotropy (Bersanelli et~al. 1997).  
However, for the purpose of beam reconstruction it is not necessary to have 
a detailed knowledge
of the planet emission. It only matters that the source is 
stable and sufficiently bright to be
detectable even when the source is far from the beam axis. 
This requirement is crucial to sample the antenna
beam response at different angles.
For this reason we will consider here only 
Jupiter and Saturn, which are the brightest of the 
external planets. On the basis of the published data,  we will assume hereafter that Jupiter and
Saturn have, at 30~GHz,   brightness temperatures  of $T^{(b)}_{jup}=152$~K and
$T^{(b)}_{sat}=133$~K, respectively.

\subsubsection{Inner bodies of the Solar System}

The amount of straylight contaminations from the inner bodies of the Solar System can be
easily quantified. We restrict ourselves to the Sun, Earth and Moon, because of their
high intrinsic temperatures and because of  the solid angles subtended from L2.  We
will use 
$T^{(th)}_{sun}=6000$~K, $T^{(th)}_{earth}=300$~K and  $T^{(th)}_{moon}=250$~K as the
temperatures associated with the millimetric thermal emission of Sun, Earth and Moon,
respectively.

\subsection{{\sc Planck} observations of the external planets }

We use the {\sc Planck} flight simulator
in order to model the transit of the planets in the {\sc Planck} field of view.  
In particular, 
Jupiter and Saturn will be observed twice in about a year.  
The solid angle of the external planets as seen by
{\sc Planck} is very small compared to the beam size. 
Thus, the {\sc Planck} observations of  the Jupiter (Saturn) will yield
\begin{equation} T_{30 GHz} [{\hat
\gamma}(t)] \simeq {T^{(b)} \pi (R/d)^2 ~J[{\hat\gamma(t)}-\hat b ]
\over 
\int_{4\pi} J({\hat \gamma})~ d \Omega }  \, .
\end{equation} 
In this equation $T_{30 GHz}$ is the observed Jupiter (Saturn) brightness temperatures @ 30
GHz; 
$T^{(b)}$ ,
 $R$
 and $d$ represent the  intrinsic Jupiter (Saturn)  brightness
temperature,  radius and distance, respectively;  $J$ is the antenna
response and ${\hat \gamma}(t)$  identifies the angular position 
of the planet as seen by {\sc Planck}, 
the time dependence  being fixed by the scanning strategy. 
\section{In-flight recovery of the main beam pattern}
\label{beam_rec}

\subsection*{}
\label{minimum}

The {\sc Planck} Time Ordered Data (TOD) are affected by instrumental noise. 
Therefore, our capability to recover the main beam pattern rests on the possibility to 
clearly detect a
bright source (e.g., Jupiter), even when significantly far from
the beam axis.  A proper description of the {\sc Planck}-LFI 
instrumental noise should in
principle include a $1/f$ contribution (see, e.g., Bersanelli et~al. 1996, 
Seiffert et~al. 1997). The
knee-frequency of the $1/f$ noise is expected to be comparable with the spinning
frequency. 
However, it has been shown that destriping algorithms can
very efficiently remove this low frequency noise component even under more 
pessimistic conditions (see, e.g., Maino et~al. 1999 and references therein) 
and return a TOD that 
we will assume, accordingly with the goals of this paper, white noise dominated. 
So, in what follows
we will model the TOD noise component as pure white noise, with the {\sc Planck}
goal sensitivities discussed by Bersanelli et~al. 1999 
(private reference, 
{\sc Planck} Low Frequency Instrument,
Instrument Science Verification Review,
October 1999, LFI Design Report).
In principle, the signal fluctuations introduced by CMB and foreground
anisotropies behave as a noise source in this context. However,  
since they can be accurately subtracted from the TOD by using the {\sc Planck}
final maps, we neglect them in what follows. We adopt here a simple scanning strategy
with $\alpha = 90^\circ$.
In the simulations presented in Sect.~3.1, 
we oversample each scan circle every $\simeq 5'$
(i.e. roughly 6 points per FWHM @ $30 GHz$) and shift the spin axis 
by $5'$ every two hours.
After simulating the Jupiter and Saturn transits we extract 
from the time ordered scans a few
($\sim 100$) chunks comprising the source transit. Since the source is 
pointlike,
these chunks give, when displayed one after the other and having taken into account
the small variations of planet distance in the different samplings,
a 2-D plot of the
beam profile, $\simeq 8.3^\circ \times 8.3^\circ$ wide.

In Fig.~2 we show
the  expected signal from Jupiter as seen along the scan circle which crosses the
source at the maximum (top panel) and as seen along an arc orthogonal 
to this circle (bottom panel).  
The signal to noise can be improved, wrt the case of a single 
receiver and transit, by considering that two LFI receivers are coupled to the
same optical beam, that there are two 30~GHz beams with the same optical properties
and that two (three) transits of both Jupiter and Saturn are expected for a one year 
(for a $14\div15$ months) mission.
This obviously
increases the signal to noise ratio by a factor of $2\sqrt{2}$ ($2\sqrt{3}$).  
As a result, @ 30 GHz, the shape of the main beam can be recovered down to
$-(25 \div 32.5)$~dB, i.e. at about $(3.5 \div 4) \sigma$.  

\begin{figure}
\resizebox{\hsize}{!}{\includegraphics{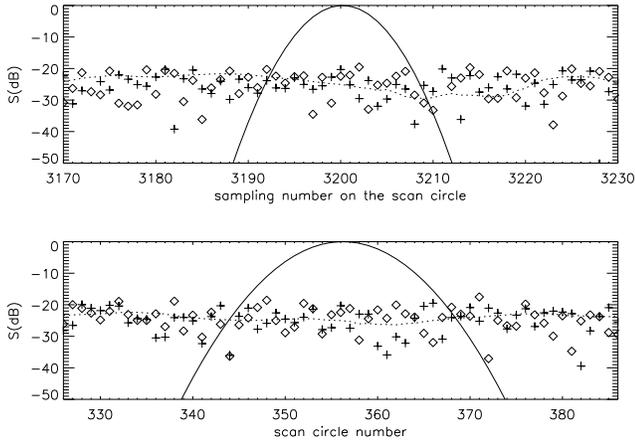}}
\caption{Top panel: signals of different components along the scan circle
with the maximum of Jupiter contribution during its first transit.
Solid line: Jupiter signal;
crosses: white noise; diamonds: $1/f$ noise coupled to white noise;
dotted line: signal from CMB and extragalactic source fluctuations
and Galaxy emission modelled according to Burigana et~al. 2000a and
references therein.
Bottom panel: the same as in the top panel, but considering the signal
at the same scan position, where Jupiter signal is maximum, for different scan circles.
According to the simulation parameters, by multiplying the scan circle number
or the sampling number on the scan circle by 5
we have respectively the angular displacement between different scan circles
and (approximately, owing to the off-axis beam position) the
angular displacement between different samplings along the scan circle,
expressed in arcmin.
We consider here the case of an elliptical beam with $r=1.3$, clearly visible
in the different spread of Jupiter signal in the two panels.
Note that the signal to noise ratio is larger than unit up to
$-(20 \div 25)$~dB, i.e. at about $(3 \div 3.5) \sigma$.  
[Signals in dB normalized to the maximum Jupiter signal at its first transit].
}
\label{limits}
\end{figure}

\subsection{Recovery of the main beam parameters}
\label{parameters}

We consider both a symmetric and an elliptical beam, with ellipticity ratio 
$r=1.3$. The numerical values of the beam parameters are shown in
Table~1. We use Eq.(1) to model the antenna response. We fit the
beam shape theoretical parameters to the 2D plot of the beam 
response obtained as
mentioned at the end of the previous section.

The~results~of~the~fits~are~shown~in~Table~2. 
We fit also an additional parameter, $r_k$,
related to the planet brightness temperature and to 
the average distance
of the planet from the spacecraft,  $\langle d  \rangle$, 
for the points considered in the fit: 
$r_k = \pi (R/\langle d  \rangle)^2 T^{(b)} / \int_{4\pi} J({\hat \gamma})~ d \Omega \,$.
We recover the full set of parameters with very high accuracy
(Burigana et~al. 2000c), the $\chi^2 /\,
\mathrm{DOF}$ being always very close (to better than $1\%$) to the
unity value. It is obviously more efficient to recover the beam pattern 
parameters using Jupiter rather than Saturn, simply because Jupiter is
brighter.

An interesting byproduct of the beam fitting procedure is the possibility
of estimating the planet's antenna temperature, or more
properly the product $T^{(b)} R^2$ relevant for the planet's
emission at the considered frequency. This quantity, as
previously stated, is a poorly known quantity at LFI's
frequencies. By considering Eq.\
(\ref{par_beam}) it is clear that the latter temperature, $T^{(b)}$
is related to the normalization of the bivariate which is in turn very
well constrained by our fit. The ability to estimate $T^{(b)}$ then
rests upon the overall calibration  accuracy\footnote{We want to stress
that the calibration of the TOD is not needed in order to reconstruct the
other beam parameters.} and on the knowledge of the total antenna beam
integral. Calibration for {\sc Planck} will be provided by continuous
observation of the CMB dipole signature and its modulation
introduced by the spacecraft motion (Bersanelli et~al. 1997)  and is expected
to be accurate to within 1\%. The total integral of the antenna pattern
poses a more serious problem, as the contribution of the far
side lobes goes undetected when using a celestial source. However,
optical calculations (de Maagt 1998) show that the contribution to the
antenna pattern coming from outside the  main lobe is expected to be
$< (2 \div 3) \%$ of the total. 
The uncertainty on the latter figure may then dominate and
pratically  set the accuracy on the LFI  estimate of $T^{(b)}$;
therefore, even a poor knowledge of this contribution
(e.g., with an accuracy of $\sim 30\%$) allow to reach a $\simeq 1\%$ level 
of accuracy in the measurement of $T^{(b)}$ (Burigana et~al. 2000c).

\label{problems}

So far no systematic effect has been included in the simulations. One of the
main goals of the {\sc Planck} mission is to limit the contamination arising
from systematics at few $\mu\mathrm{K}$ level. 
However, systematics affecting the spacecraft pointing and rotation, which
are not expected to significantly degrade the sky temperature estimation,
may in principle turn out to be harmful when reconstructing the beam shape.
To investigate this possibility we have extended our simulation to include
these effects. 

\subsubsection{Pointing uncertainty}

A potential problem for in-flight beam reconstruction is posed
by the telescope pointing uncertainty, i.e. the limited accuracy in
determining the spacecraft's effective spin axis. However, it is a
{\sc Planck} requirement that such an uncertainty be less than $1'$. 
This is
why, to simulate this effect  we introduce a
pointing uncertainty 
drawn from uniform  distribution of values between 0 and 1. Since the
spin axis direction is not expected to significantly change within the same
scan, the uncertainties only arising when the spacecraft is repointed, we
keep the perturbation constant along the scan.

The impact of pointing uncertainty in the elliptical beam parameter recovery for the
first transit of Jupiter and Saturn is also shown in Table~2.  
Clearly, the higher signal to
noise ratio in case of Jupiter  makes the results more sensitive to this kind of
systematics. This is evident from the larger degradation of the beam parameter
recovery. Although the degradations in the recovered parameters are well above
the corresponding quoted statistical errors, we do not 
find a particularly critical
effect on the various parameters, except for the parameter
$r-1$,  affected by an error of 10\% by using Jupiter, for  the adopted level of
pointing uncertainty.  On the other hand, the analysis of the $\chi^2 /\,
\mathrm{DOF}$ ($\simeq 1.3$) clearly flags this problem. In this circumstance,
it may be then particularly safe  to take advantage of the Saturn transits, that,
although intrinsically noisier, are significantly less affected by this kind of
systematics. We can then exploit the Jupiter transits to check the fit results where
the signal to noise ratio from Saturn  transits decreases.

The reported values of $\chi^2 /\, \mathrm{DOF}$ can be understood on the
basis  of simple considerations. The derivative of a Gaussian response, 
$J={\rm exp}[-(\theta/\sigma)^2/2]$, with respect to the angle $\theta$ from the 
beam center at $\theta \sim \sigma$ is given by $dJ/J \sim -d\theta/\sigma$;  for
Jupiter (Saturn), the signal at $\theta \sim \sigma$ is
$\sim 22$~mK ($\sim 5$~mK) for a beam with a FWHM~$\sim 33'$. Then, a
pointing error $d\theta \sim 1'$ about $\theta \sim \sigma$  implies a signal
change of $\sim 1.6$~mK ($\sim 0.35$~mK ).  Let's assume such value of error in a
squared region about the beam center  with a 2$\sigma$ side, containing
$\sim 32$ samplings in the current  simulations. The fit uses $\sim 10^4$
samplings with a white noise sensitivity of $\sim 0.17$~mK.  We find then a
$\chi^2 /\, \mathrm{DOF}$ of $(10^4 \times 0.17^2 + 32 \times 1.6^2) / (10^4
\times 0.17^2) \sim 1.3$ for  Jupiter (and analogously $\sim 1.01$ for Saturn),  in
good agreement with the values reported in Table~2.

\subsubsection{Spacecraft rotation}

We have also verified the effect
the spacecraft
rotation during single sample integration.
Strictly speaking, the signal inside each sampling time changes according to the beam
response. We can take this effect into account by oversampling and then
averaging the simulated data inside each sampling time and by equivalently
implementing in the fit procedure the recovery of the intrinsic (i.e. non affected by
rotation)  optical beam parameter. We have here implemented the impact of the
satellite rotation only in the data flow generation, in order to estimate the error 
introduced in the beam parameter recovery by a simple fit procedure  that does not
take into account this effect. The results are again reported in Table~2 for the first
transit of Jupiter and Saturn  in the case of an elliptical beam. As expected, the effect
on beam position and inclination is negligible. Since in the present test the minor
axis is taken along the scan circle direction, the beam is reconstructed with a smaller
ellipticity  and a larger FWHM with respect to the input values, with no indication
of a worsening of the $\chi^2 /\,
\mathrm{DOF}$. Clearly, in the general case, the final effect will depend on a
combination of beam ellipticity and inclination with respect to the sky scan
direction.  
The beam parameters recovered by neglecting the spacecraft spinning
within the sampling time in the fit procedure may also be seen as characterizing the
``effective'' properties of  the beam  in presence of spacecraft rotation. Again, the
``effective'' values of $\sigma$ reported in Table~2  are in agreement with simple
analytical considerations. In fact, 
the effective resolution which takes  into account the intrinsic beam
optical resolution and the beam smearing introduced by the  satellite rotation is well
approximated by $\sigma_{eff} = \sqrt{ \sigma^2 +
\theta_s^2/12  }$ where $\theta_s$ is the angle ($\simeq 5'$ for the present
simulations)  in the sky described by the beam axis during the sampling time.

\begin{table*}[t]
\caption[InpVal]{Input parameters of symmetric and elliptical beam for the
considered planet transits.}
\begin{tabular}{|l||c|c|c|c|c|c|c|c|}
\hline
\multicolumn{9}{|l|}{\em Input values}\\ \hline
               & \multicolumn{2}{|c|}{$\epsilon$} & $\sigma$ 
&\multicolumn{2}{|c|}{$r$} & $r_k$\footnotesize{$\;^\ddagger$} & 
$x^*_T\cdot10^{2}$ & $y^*_T\cdot10^{2}$ \\ Event      &
\multicolumn{2}{|c|}{(deg)}   &   (arcmin) &\multicolumn{2}{|c|}{$-
$}    & (mK)      &  $-$    & $-$ \\ 
               & {\em circ.} & {\em ellipt.} & & {\em circ.} & {\em ellipt.} &          &       &
\\\hline  Jupiter (I)    &           &            & &            &            & $35.8297$  &      
&\\\cline{1-1}\cline{7-7} Jupiter (II)    &          &            & &            &           &
$35.4743$ &        &\\\cline{1-1}\cline{7-7} Saturn (I)      & \rb{$-$}  &   \rb{$0$} &
\rb{$14.01381$}&\rb{$1$}&\rb{$1.3$}&  
$7.8543$  & \rb{$-5.76035$}&\rb{$7.91971$}\\\cline{1-1}\cline{7-7} Saturn (II)   
&           &             & &            &            & $7.8229$   &       &\\\hline
\multicolumn{9}{l}{\footnotesize{$^\ddagger$ The value $r_k$ depends on the 
distance between the spacecraft and the planet which slightly varies}}\\
\multicolumn{9}{l}{\footnotesize{between different pointing events.}}  
\end{tabular}
\end{table*}
\begin{table*}[t]
\caption[Fits]{Recovery of the beam parameters from the considered planet transits
in presence of pure white noise and taking also into account two kinds of
systematical effects. Circular, elliptical and simulated beams as well as
different values of the boresight angle $\alpha$ are considered.}
\begin{tabular}{|l||c|c|c|c|c|c|c|}

\hline
               & $\epsilon$ & $\sigma$ & $r$ & $r_k$        & $x^*_T\cdot10^{2}$ & 
$y^*_T\cdot10^{2}$ & $\chi^2 /\, \mathrm{DOF}$\\
\rb{Event}      &  (deg)   &   (arcmin) &   $-$    & (mK)      &  $-$    & $-$ & 
$-$    \\ \hline
\multicolumn{8}{|l|}{\em Circular beam}\\ \hline
                &          & $14.0071$   & $1.0039$    & $35.877$   &$-5.7602$     &$7.9192$ &
\\
\rb{Jupiter (I)}  & \rb{$-$} &$\pm 0.0089$ & $\pm 0.0012$ & $\pm 0.032$ &
$\pm  0.0004$&$\pm 0.0004$ & \rb{$0.994$}\\
\hline
                 &          & $14.0126$  & $1.0010$    &  $35.486$  &$-5.7603$     &$7.9200$ & 
\\
\rb{Jupiter (II)} & \rb{$-$}& $\pm 0.0092$ & $\pm 0.0013$ & $\pm 0.033$ &
$\pm  0.0004$&$\pm 0.0004$ & \rb{$0.997$}\\
\hline
                &          &  $13.996 $   & $1.0067$    & $7.839$   &$-5.7659$     &$7.9204$ &
\\   
\rb{Saturn (I)  } & \rb{$-$}& $\pm 0.042$ & $\pm 0.0061$    & $\pm 0.033$ &
$\pm  0.0020$&$\pm 0.0015$ &\rb{$0.998$}\\ \hline
                &          &  $14.034$   & $1.0078$    & $7.797$   &$-5.7637$     &$7.9209$ &
\\
\rb{Saturn (II) } & \rb{$-$}& $\pm 0.042$ & $\pm 0.0061$    & $\pm 0.033$ &
$\pm  0.0020$&$\pm 0.0015$ & \rb{$1.007$}\\ \hline
\multicolumn{8}{|l|}{\em Elliptical beam}\\ \hline
                & $-0.0059$   & $14.0095  $   & $1.3023$    & $35.876$   &$- 5.7602$   
&$7.9193$ & \\
\rb{Jupiter (I)} &$\pm 0.0024$ & $\pm 0.0088$ & $\pm 0.0016$ & $\pm 0.032$ & 
$\pm 0.0004$&$\pm 0.0003$ & \rb{$0.995$}\\
\hline
                & $0.0011$   & $14.0102  $   & $1.2993$    & $35.491$   &$- 5.7603$   
&$7.9199$ &  \\
\rb{Jupiter (II)} &$\pm 0.0025$ & $\pm 0.0092$ & $\pm 0.0017$ & $\pm 0.033$ & 
$\pm 0.0004$&$\pm 0.0003$ & \rb{$0.997$} \\
\hline
                &  $-0.007$    &  $14.004 $   & $1.3055$    & $7.837$   &$- 5.7661$   
&$7.9199$ &  \\
\rb{Saturn (I) } & $\pm 0.011$ & $\pm 0.042$ & $\pm 0.0078$    & $\pm 0.033$ & 
$\pm 0.0020$&$\pm 0.0015$ & \rb{$0.999$}\\
\hline
                &  $0.009$    &  $14.041 $   & $1.3073$    & $7.796$   &$- 5.7645$    &$7.9207$
&  \\
\rb{Saturn (II) } & $\pm 0.011$ & $\pm 0.042$ & $\pm 0.0078$    & $\pm 0.033$ & 
$\pm 0.0020$&$\pm 0.0015$ &\rb{$1.007$} \\
\hline
\multicolumn{8}{|l|}{\em Elliptical beam: effect of $1'$ pointing error}\\ 
\hline
                & $0.0288$   & $13.7895  $   & $1.3424$    & $35.884$   &$- 5.7576$   
&$7.9227$ & \\
\rb{Jupiter (I)} &$\pm 0.0025$ & $\pm 0.0089$ & $\pm 0.0017$ & $\pm 0.032$ & 
$\pm 0.0004$&$\pm 0.0003$ & \rb{$1.329$}\\
\hline
                &  $-0.005$    &  $14.004 $   & $1.3060$    & $7.819$   &$- 5.7589$   
&$7.9118$ & \\
\rb{Saturn (I) } & $\pm 0.011$ & $\pm 0.043$ & $\pm 0.0078$    & $\pm 0.033$ & 
$\pm 0.0020$&$\pm 0.0015$ & \rb{$1.011$} \\
\hline
\multicolumn{8}{|l|}{\em Elliptical beam: effect of neglecting spacecraft 
motion}\\ \hline
                & $0.00367$   & $14.0697$  & $1.2904$    & $35.584$   &$-5.7602$    
&$7.9198$ & \\
\rb{Jupiter (I)} &$\pm 0.0025$ & $\pm 0.0089$ & $\pm 0.0016$ & $\pm 0.032$ & 
$\pm 0.0004$&$\pm 0.0003$ & \rb{$0.999$} \\
\hline
                &  $-0.007$    &  $14.053 $   & $1.2964$    & $7.782$   &$- 5.7661$   
&$7.9200$ &  \\
\rb{Saturn (I) } & $\pm 0.012$ & $\pm 0.043$ & $\pm 0.0078$    & $\pm 0.033$ & 
$\pm 0.0020$&$\pm 0.0015$ &\rb{$0.999$}  \\
\hline
\multicolumn{8}{|l|}{\em Alcatel case1 beam: fit with circular beam -- 
realistic simple scanning strategy, $\alpha = 80^{\circ}$}\\ \hline
                &              & $14.9583$  &          & $31.106$   &$-5.5751$    
&$7.2991$ & \\
\rb{Jupiter (I)} & \rb{$-$} & $\pm 0.0089$  & \rb{$-$} & $\pm 0.026$ & 
$\pm 0.0003$&$\pm 0.0004$ & \rb{$8.094$} \\
\hline
\multicolumn{8}{|l|}{\em Alcatel case1 beam: fit with elliptical beam -- 
realistic simple scanning strategy, $\alpha = 80^{\circ}$}\\ \hline
                & $20.97$   & $15.0102$  & $1.3602$    & $31.360$   &$-5.5756$    
&$7.3001$ & \\
\rb{Jupiter (I)} &$\pm 0.11$ & $\pm 0.0087$ & $\pm 0.0016$ & $\pm 0.026$ & 
$\pm 0.0003$&$\pm 0.0004$ & \rb{$1.232$} \\
\hline
\multicolumn{8}{|l|}{\em Above elliptical beam: fit with elliptical beam  -- 
baseline simple scanning strategy, $\alpha = 85^{\circ}$}\\ \hline
                & $21.01$   & $15.0209$  & $1.3590$    & $32.241$   &$-5.5792$    
&$7.2794$ & \\
\rb{Jupiter (I)} &$\pm 0.10$ & $\pm 0.0083$ & $\pm 0.0015$ & $\pm 0.025$ & 
$\pm 0.0003$&$\pm 0.0004$ & \rb{$0.997$} \\
\hline\hline
\end{tabular}
\end{table*}

\subsection{Recovery of the main beam shape}
\label{shape}

We consider here the capability to reconstruct in flight the 
detailed shape of the main beam by using Jupiter transits. 
We apply our method to a beam @ 30~GHz simulated through the GRASP8 code
for the Alcatel case1 telescope configuration (see Fig.~3), an aplanatic solution
like those suggested by Mandolesi et~al. 2000b in order to minimize
the coma distortion and render the main beam shapes close to ellipses.
In this case we adopt a simple scanning strategy with $\alpha = 80^\circ$, the same
scan angle considered in the Alcatel case1 design, shift the spin axis of $2.5'$
every hour and oversample each scan circle every $\simeq 11'$ 
(i.e. rougly 3 points per FWHM @ $30 GHz$).

\begin{figure}
\resizebox{\hsize}{!}{\includegraphics{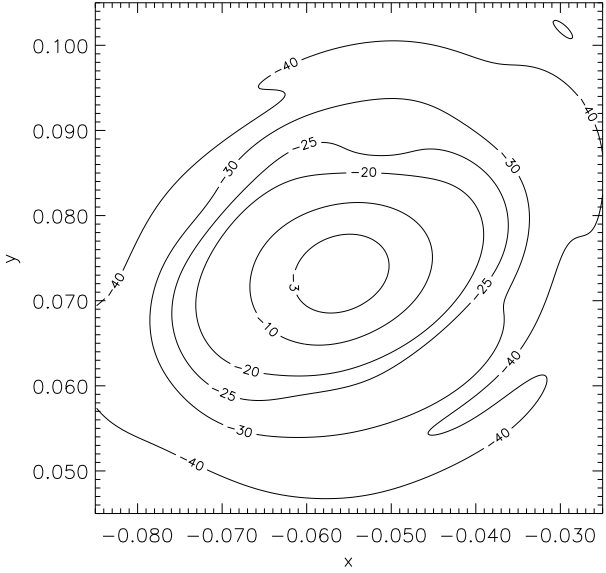}}
\caption{Contour plot (in dB) of one of the two beams @ 30 GHz as 
simulated through the GRASP8 code for the Alcatel case1 telescope 
configuration.}
\label{input}
\end{figure}

By inverting Eq.\
(\ref{par_beam}) we derive the beam pattern shape. 
Fig.~4 show the results obtained @ 30~GHz by considering the sensitivity 
of the two radiometers coupled to a single beam and a single Jupiter transit:
the main beam shape can be directly recovered down to $\simeq -20$~dB 
with good accuracy.
The result clearly improves by adding three Jupiter transits
and taking into account the possibility of averaging the recovered shapes 
of the two equivalent beams at the same frequency, as shown in Fig.~5.

\begin{figure}
\resizebox{\hsize}{!}{\includegraphics{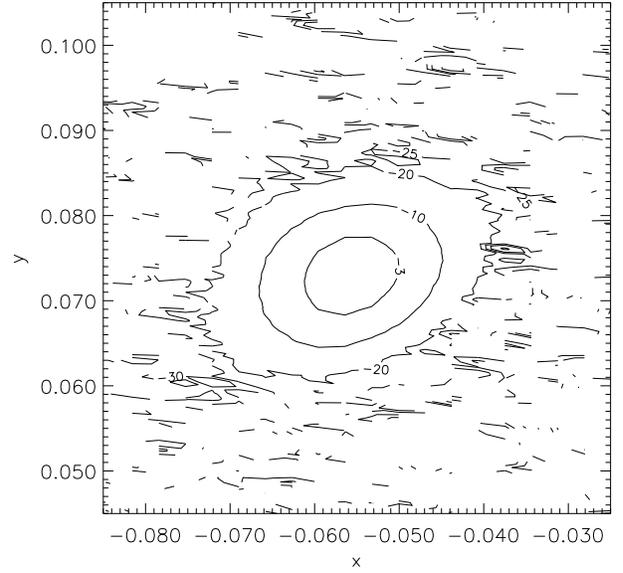}}
\caption{Reconstruction of the beam of Fig.~3 by using a single Jupiter transit 
and the sensitivity of the two radiometers coupled to the beam.}
\label{output}
\end{figure}
\begin{figure}
\resizebox{\hsize}{!}{\includegraphics{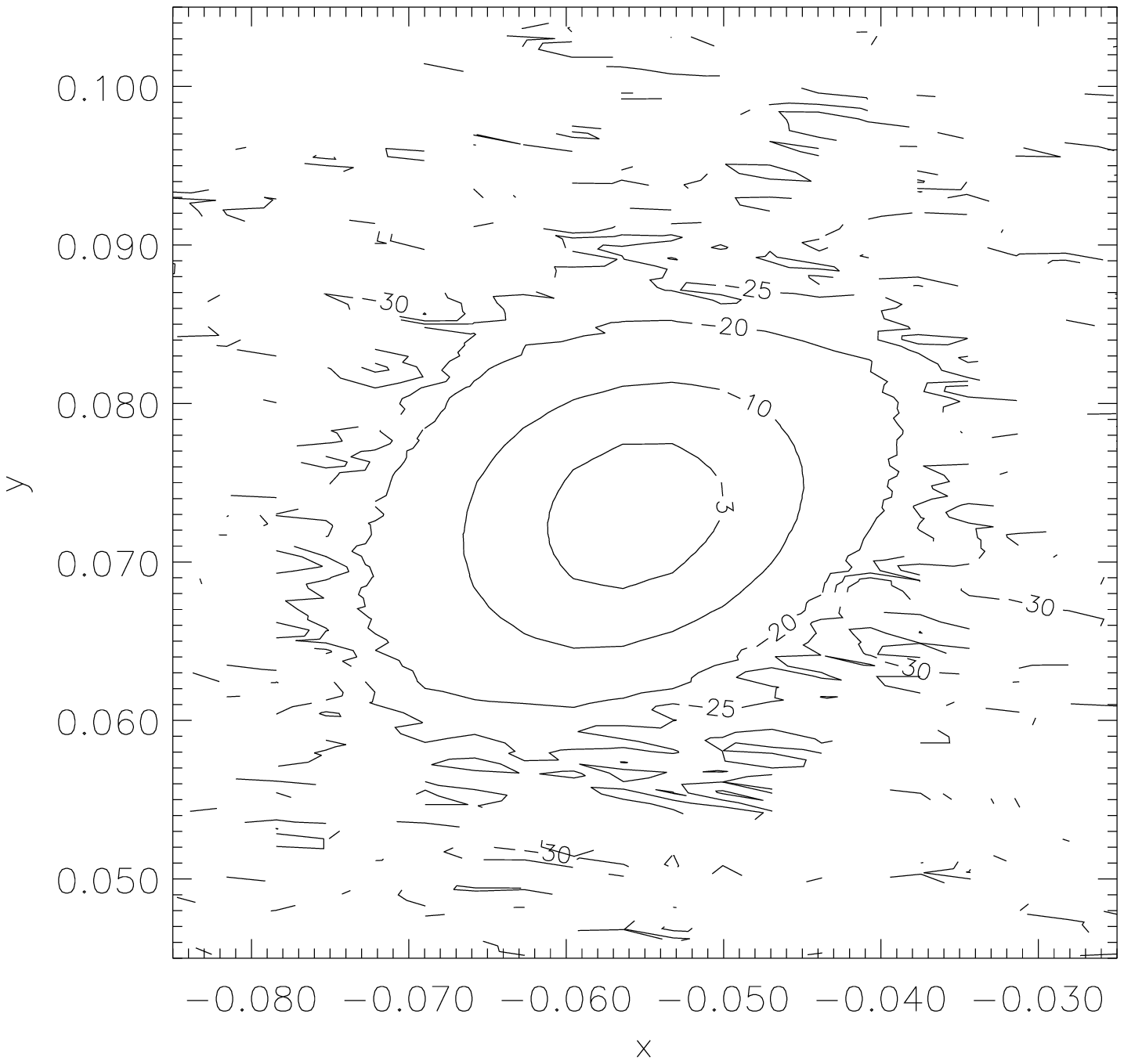}}
\caption{Reconstruction of the beam of Fig.~3 
by using three Jupiter transits and two radiometers
per beam and by taking advantage from the identical optical properties
of the two 30~GHz beams.}
\label{outputgood}
\end{figure}

By fitting the TOD with the method described in Sect.~3.1 we can derive the
beam parameters under the assumption of a circular or bivariate Gaussian  
shape. The results are reported in Table~2.
We show also respectively in Fig.s~6 and 7 the comparison between the input beam shape
and the recovered one for the case of circular and bivariate Gaussian  
approximation.
As evident, the bivariate approximation represents a significant 
improvement with respect to the circular one, as indicated by the 
value $\chi^2 /\, \mathrm{DOF}$, respectively 1.232 and 8.094.
The agreement between the bivariate approximation and 
the simulated beam results quite good down to $\simeq -25$~dB.
At lower response levels the Gaussian behavior begins to 
significantly underestimate the beam response,
even in the bivariate approximation.

\begin{figure}
\resizebox{\hsize}{!}{\includegraphics{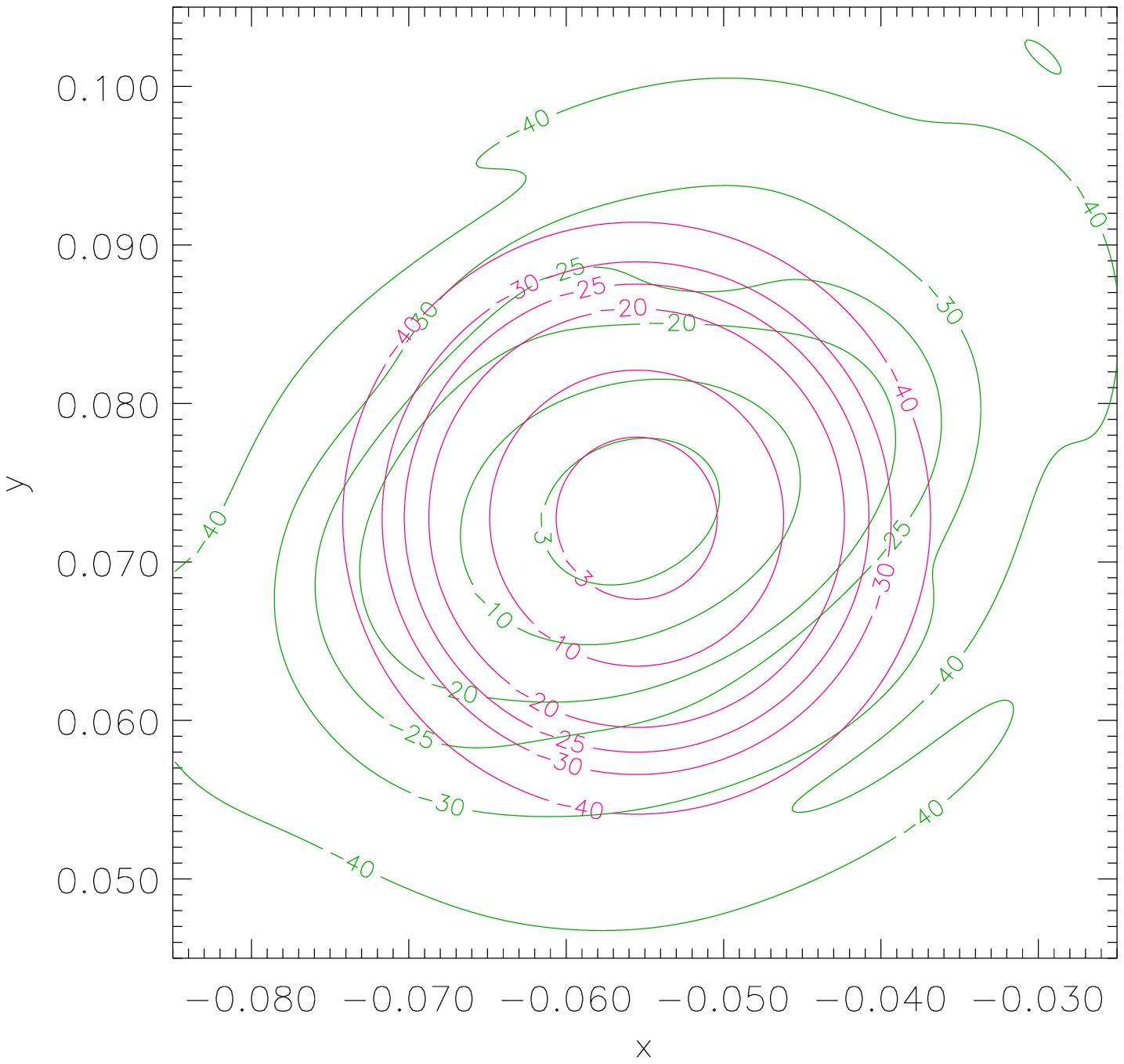}}
\caption{Comparison between the simulated beam and its approximation
in terms of circular Gaussian beam (see also Table~2).}
\label{sym}
\end{figure}
\begin{figure}
\resizebox{\hsize}{!}{\includegraphics{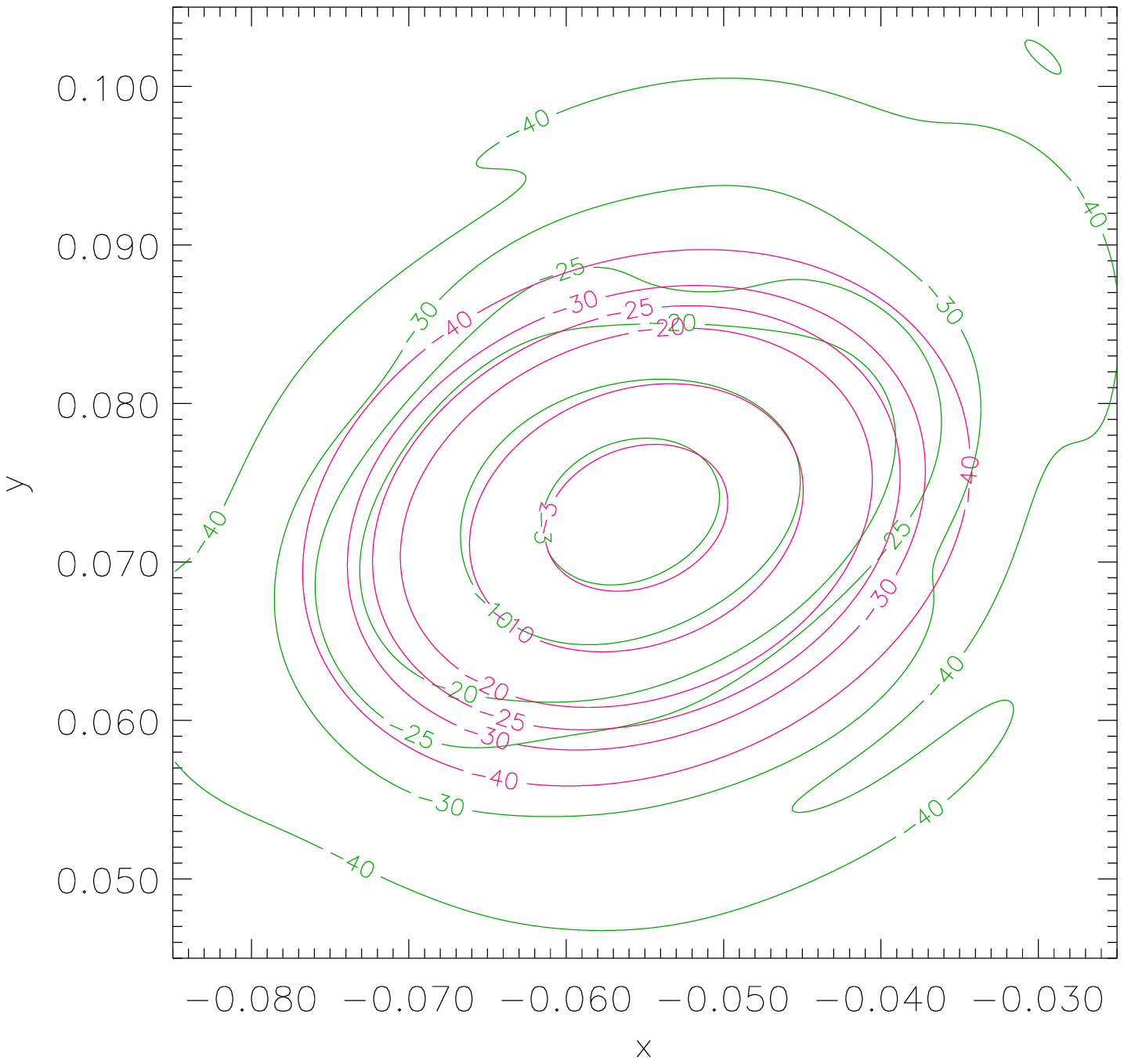}}
\caption{Comparison between the simulated beam and its approximation
in terms of elliptical Gaussian beam (see also Table~2).}
\label{sym_ell}
\end{figure}

Finally, we simulate Jupiter transits as seen by an elliptical beam with the
same parameters derived from the fit of the simulated beam, but  
in the case of the current simple baseline scanning strategy defined by
$\alpha = 85^\circ$, spin axis shift of $2.5'$
every hour and by approximately 3 samplings per FWHM (i.e. one point every 
$\simeq 11'$ along a scan circle @ 30~GHz).
As expected, the beam parameters are recovered with an accuracy
essentially unchanged with respect to the results of Sect.~3.1
(see the last row of Table~2), in spite 
of the less regular grid in the $x_T - y_T$ plane, less refined 
(by a factor $\simeq 4$) along the scan direction
than along that of the spin axis repointing.

\subsection{Circular versus elliptical approximation of the beam shape}

A crucial point for CMB anisotropy experiments is 
to estimate the impact of the quality 
of the main beam reconstruction on the data analysis and, ultimately,
on the sky maps that can be obtained and on the related science.
An accurate discussion of this problem is not the aim of this work.
Nevertheless, we want address here this argument.
By including also the Galaxy emission, 
only at low Galactic latitudes and at the lowest LFI frequencies
the impact of elliptical main beam 
distortions significantly increases (by a factor $\simeq 3$) with respect to 
the case of a pure CMB fluctuation sky (Burigana et~al. 1998) whereas
only very small effects are added by the combined effect of 
realistic main beam distortions and extragalactic 
source fluctuations (Burigana et~al. 2000a). In addition,
the main beam distortion impact moderately increases with the FWHM
(Burigana et~al. 1998) for the LFI resolution range.
Therefore, we will consider here for simplicity a pure CMB fluctuation sky 
in a reference case of a standard CDM model approximately
COBE/DMR normalized and consider for the present analysis the 
30~GHz channel.
In Fig.~8 we report the difference between the signals obtained
by convolving the sky with the bivariate main beam and 
with the circular main beam 
by assuming in each case the corresponding set of parameters
that fits the simulated beam (see Table~2). 
The rms of the temperature differences is $\simeq 4.1 \mu$K.
A similar comparison but between the circular and the simulated beam
(see Fig.~9) gives a very similar rms value, $\simeq 4.7 \mu$K.
This suggests that the ellipticity is the most relevant 
feature of the main beam shape.
In fact, the comparison between the convolutions with the bivariate Gaussian
beam and the simulated one (see Fig.~10) gives a rms temperature 
difference of $\simeq 1.7 \mu$K, a factor $\simeq 3$ smaller than  
that obtained by approximating the beam with a symmetric Gaussian shape.
\begin{figure}
\resizebox{\hsize}{!}{\includegraphics{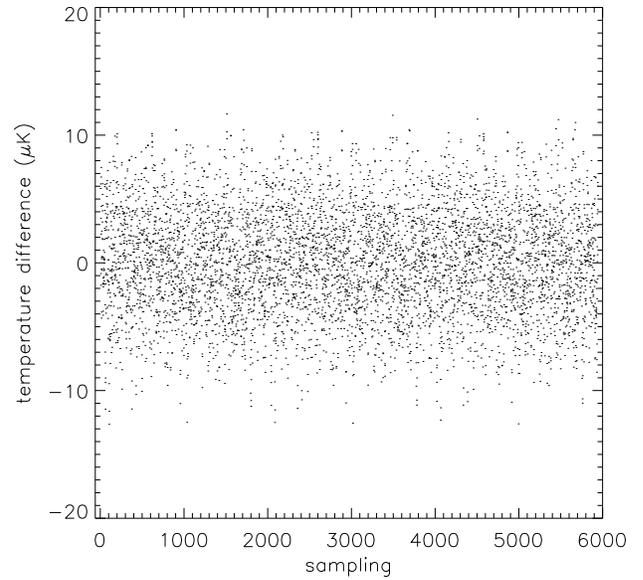}}
\caption{Difference between the TOD from the elliptical beam and 
the circular beam that fit the input beam
for three scan circles.}
\label{te_ts}
\end{figure}
\begin{figure}
\resizebox{\hsize}{!}{\includegraphics{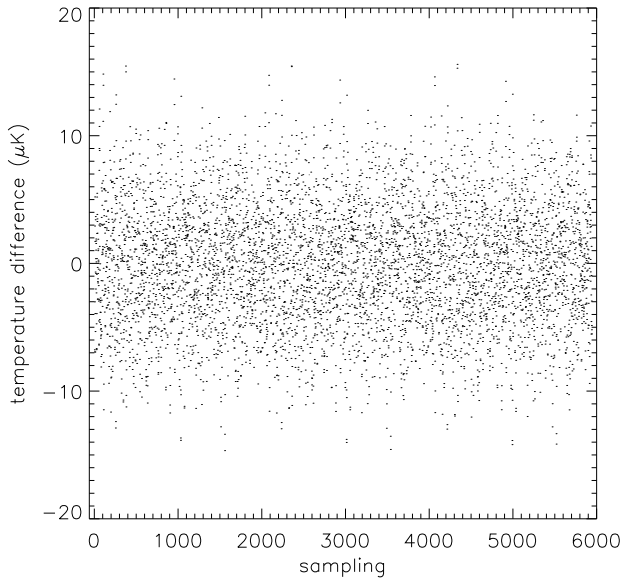}}
\caption{Difference between the TOD from the circular beam and the simulated beam
for the same three scan circles of Fig.~8.}
\label{ts_tt}
\end{figure}
\begin{figure}
\resizebox{\hsize}{!}{\includegraphics{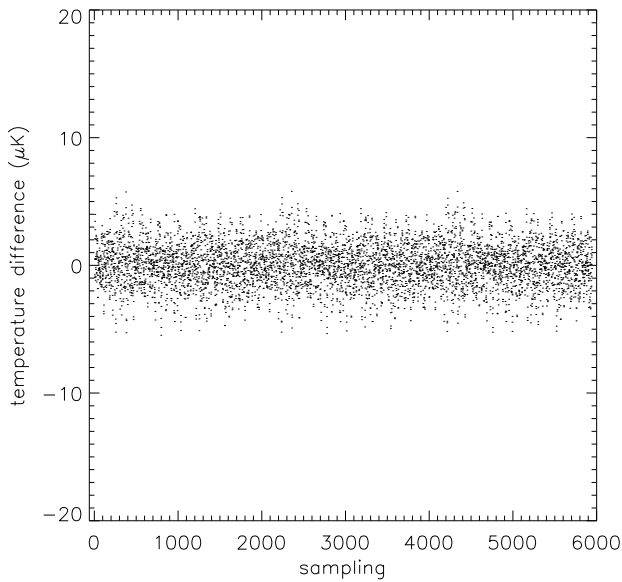}}
\caption{Difference between the TOD from the elliptical beam and the simulated beam
for the same three scan circles of Fig.~8. 
By comparing this figure with Fig.~9
the improvement with respect to the
circular approximation of the main beam is evident.} 
\label{te_tt}
\end{figure}

Of course, when the {\sc Planck} optical design will be settled and the 
main beam patterns computed through optical simulation codes 
it would be possible to search for
analytical descriptions of main beam shapes that might improve the 
bivariate Gaussian approximation. Anyway, we have proved here
that this representation allow to reach the $\simeq \mu$K accuracy level 
in the treatment of {\sc Planck}-LFI TOD's.
\section{Straylight from Moon, Earth and Sun}
\label{internal_straylight}

By using the same method described in Sect.~2 and the full antenna pattern
shown in Fig.~1 (de Maagt et~al. 1998), 
we are able to quantify the impact of the internal Solar System
objects to the {\sc Planck}-LFI observations. 
The Moon, the Earth and the Sun
are the only internal bodies that may introduce appreciable straylight
contamination. 
The solid angle of these objects, although not negligible,
is very small compared to the angular scale for which significant variations
of the far pattern response occur; this make the use of Eq.~(2)
accurate enough for the present purposes.

Of course, the level of straylight
contamination from these bodies depends also on the effective scanning strategy.
We studied a Lissajous orbit around L2 for the {\sc Planck} surveyor (Bersanelli et~al. 1996) 
by considering $case \, i)$ the simple  scanning strategy with the spin axis always
parallel to the Sun--spacecraft  direction  and $case \, ii)$ a precession motion of the
satellite spin axis like that discussed in Sect.~2.1.  [For this straylight analysis we
assume $\alpha = 80^{\circ}$, the value adopted in the optical computation by
de~Maagt et~al. 1998]. 

By coadding the TOD computed as described in Sect.~2 we have produced
nearly full sky maps
of averaged straylight contamination (see Burigana et~al. 2000b for further 
details). [We adopt here the HEALPix pixelisation by G\'orski et al.~1998].
We find that in any case this effect is very small, with the maximum
contamination level always below $0.15~\mu\mathrm{K}$.
Although small, this effect has to be considered with caution. In fact,
this straylight contamination is produced by the very low
response antenna in the ``antispillover'' region,  where an accurate 
antenna response
measurement is extremely difficult.
\section{Discussion and conclusions}
\label{conclusions}

We implemented the {\sc Planck} flight simulator (Burigana et~al. 1997, 1998, 
Maino et~al. 1999) to properly discuss the impact
of the Solar System main bodies on the {\sc Planck} observations. In particular, we focused on
the problem of the in-flight reconstruction of the main beam of the {\sc Planck}-LFI antenna
patterns. To do so, we simulate in details the transits of Jupiter and Saturn in the
field of view of the {\sc Planck}-LFI, 30 GHz beam. The method can be easily extended to the
other {\sc Planck} channels. 
Our analysis shows that, using Jupiter, we can recover in flight the main beam response
down to  $\sim -(25 \div 32.5)$~dB, where the signal to noise ratio 
approaches unity.  

Both symmetric and non symmetric beams have been considered; in the latter case 
we assumed a simple elliptical shape, as suggested by recent optical  simulations 
(Mandolesi et~al. 2000b),
but the method can be generalized to more refined parametrizations.

We have demonstrated that the key parameters of the 
main beam (resolution, ellipticity,
position and inclination on the plane of {\sc Planck} field of view)  can be
simultaneously recovered with high precision by fitting the planet  transit signal. Of
course, the larger signal to noise ratio of Jupiter (compared to that of Saturn)
translates in a better parameter recovery, by a factor $3 \div  5$.

After having considered the idealized case of pure white noise, we discussed 
possible degradations to the main beam reconstruction introduced 
by the effects of spacecraft pointing relevant in this context. 
Spacecraft rotation does not affect
significantly the quality of the reconstruction; on the contrary, 
bu neglecting the spacecraft rotation in the fit procedure
we derive the effective beam parameters including rotation smearing effect.
The most relevant source of contamination
is represented by the spin axis  pointing uncertainty 
which can degrade the recovery
of the beam ellipticity parameter
$r-1$ by $\simeq 10\%$ in the case of 1$'$ pointing error when Jupiter is used. In
these circumstance, clearly recognizable through the increase of the 
$\chi^2 /\, \mathrm{DOF}$, we find very advantageous to consider  Saturn, that
produce information less sensitive to this  kind of systematics.

To complete the analysis, we considered also the full recovery of a simulated beam
shape for an aplanatic configuration of the optical design recently studied 
by Alcatel. The beam shape can be reconstructed 
with good accuracy and resolution down to $- 25$~dB. Of course, somewhat lower response 
levels at larger angles from the beam centre can be measured in flight by 
relaxing the requirement of recovering the beam shape with the same
resolution necessary to accurately describe the more central regions of the main beam.

We addressed the study of the impact of the quality of the main beam reconstruction
on the treatment of LFI TOD's. We shown that a bivariate Gaussian approximation
represents a significant improvement with respect to a symmetric representation.
A rms difference of $\simeq 1.7 \mu$K is found between the TOD obtained
by convolving a CMB fluctuation sky with a simulated beam or its bivariate Gaussian
approximation as derived by the in-flight reconstruction, a result 
better by a factor three with respect to the case of the symmetric approximation.

The possibility to combine a very accurate in-flight calibration by using the  CMB
dipole (Bersanelli et al. 1997) 
and the good accuracy in the recovery of  the maximum
signal (the parameter $r_k$ in Table~2) at the planet transit,  offers a good chance 
of measuring the intrinsic planet temperatures at millimetric wavelengths with  an
accuracy at \% level, the main source of error being the uncertainty on the integrated  antenna
pattern response.  
This represents an interesting byproduct of {\sc Planck} observations.

Finally, we have shown that the effect of Sun, Earth and Moon in the far
sidelobes produces sub-$\mu\mathrm{K}$ effects independent of the details
of the scanning strategy.

To summarize, at least at 30~GHz, observation of external planets 
offers an accurate, robust and simple 
method to reconstruct in flight the main beam
properties under very general  conditions.




\bigskip
\bigskip

\noindent {\bf Acknowledgements.} It is a pleasure to thank  C.R.~Butler,
B.~Cappellini, G.~Cremonese, D.~Maino, F.~Pasian and F.~Villa for  useful discussions on {\sc Planck}
design and performances and  on planet emission properties. We gratefully
acknowledge K.M.~G\'orski and all the people involved in the realization of the
tools of HEALPix pixelisation, employed here in the straylight analysis. We gratefully thank
P.~de~Maagt and J.~Tauber for having provided us with their optical simulation
results and F.~Villa for having computed with the GRASP8 code the main beam 
pattern LFI27 for the Alcatel case1 configuration with the 
high resolution grid adopted in this study.

\bigskip
\bigskip
\bigskip


\end{document}